\documentclass[fleqn]{article}
\usepackage{bm}
\usepackage{epsfig}
\usepackage{amsmath,amsfonts,amssymb}
\usepackage{color}
\usepackage{espcrc2}
\usepackage{graphicx}

\newcommand{\jpsi}{J/\psi}
\newcommand{\psip}{\psi(2S)}
\newcommand{\psipto}{\psi(2S)\rightarrow}
\newcommand{\jpsito}{J/\psi\rightarrow}
\newcommand{\chicj}{\chi_{CJ(J=0,1,2)}}
\newcommand{\ee}{e^+e^-}
\newcommand{\mumu}{\mu^+\mu^-}
\newcommand{\pipi}{\pi^+\pi^-}
\newcommand{\kk}{K^+K^-}
\newcommand{\ppb}{p\bar{p}}
\newcommand{\pbar}{\bar{p}}
\newcommand{\gam}{\gamma}
\newcommand{\ra}{\rightarrow}
\newcommand{\llb}{\Lambda\bar{\Lambda}}

\newcommand{\lbar}{\bar{\Lambda}}

\newcommand{\ssb}{\Sigma^0\bar{\Sigma}^0}

\newcommand{\xxb}{\Xi^-\bar{\Xi}^+}

\def\Journal#1&#2&#3(#4){#1{\bf #2}, #3 (#4)}

\def\NIMA{Nucl. Inst.  and Meths. {\bf A }}

\def\PLB{Phys.  Lett.  {\bf B }}

\def\PRD{Phys.  Rev.  {\bf D }}
\def\ZPC{Zeit. Physik {\bf C }}
\def\nse{Nucl.  Sci.  Eng.  }

\def\hepnp{HEP \& NP }
\def\ijmpa{Int.  J.  Mod.  Phys.  {\bf A} }
\def\etal{{\it et al.}}

\def\bec{\begin{center}}
\def\eec{\end{center}}

\begin{document}
\title{Measurements of $\psip$ decays to octet baryon-antibaryon pairs }
\author{\begin{footnotesize}
\begin{center}
M.~Ablikim$^{1}$,              J.~Z.~Bai$^{1}$, Y.~Ban$^{12}$,
X.~Cai$^{1}$,                  H.~F.~Chen$^{17}$,
H.~S.~Chen$^{1}$,              H.~X.~Chen$^{1}$, J.~C.~Chen$^{1}$,
Jin~Chen$^{1}$,                Y.~B.~Chen$^{1}$
% ...
\\
\vspace{0.2cm}
(BES Collaboration)\\
\vspace{0.2cm}
{\it
$^{1}$ Institute of High Energy Physics, Beijing 100049, People's Republic of China\\
% ...
$^{12}$ Peking University, Beijing 100871, People's Republic of China\\
% ...
$^{17}$ University of Science and Technology of China, Hefei 230026, People's Republic of China\\
% ...
}
\end{center}
\end{footnotesize}
}
\date{\today}
\begin{abstract}
  With a sample of $14\times10^6$ $\psip$ events collected by the
  BESII detector at the Beijing Electron Positron Collider (BEPC), the
  decay channels $\psipto B_8\bar{B_8} ~(\ppb,\ \llb,\ \ssb,\ \xxb)$
  are measured, and their branching ratios are determined to be
  $(3.36\pm0.09\pm0.25)\times10^{-4}$,
  $(3.39\pm0.20\pm0.32)\times10^{-4}$,
  $(2.35\pm0.36\pm0.32)\times10^{-4}$,
  $(3.03\pm0.40\pm0.32)\times10^{-4}$, respectively. In the decay
  $\psipto\ppb$, the angular distribution parameter $\alpha$ is
  determined to be $0.85\pm0.24\pm0.04$.
\end{abstract}
\maketitle
\normalsize
\setcounter{section}{0}
\setcounter{equation}{0}
\section{ \boldmath Introduction}
The branching ratios of $\psip$ decays into octet
baryon-antibaryon pairs were measured by the BES-I and CLEOc
collaborations, and the results differ significantly, as shown in
Table~\ref{br_list}. It is therefore important to make new
measurements to help clarify these differences using the sample of
$14\times10^6$ $\psip$ events collected by BESII, which is the
world's largest $e^+ e^-$ $\psip$ sample.

\begin{table}[h!]
\caption{\label{br_list}Branching ratios of $\psipto
B_8\bar{B_8}(\times10^{-4})$.}
\begin{center}
\begin{tabular}{lll}\hline\hline
Channel          &BES-I~\cite{besi}&CLEO-c~\cite{cleo}\\
\hline
$\ppb$&$2.16\pm0.15\pm0.36$&$2.87\pm0.12\pm0.15$\\
$\llb$&$1.81\pm0.20\pm0.27$&$3.28\pm0.23\pm0.25$\\
$\ssb$&$1.2\pm0.4\pm0.4$   &$2.63\pm0.35\pm0.21$\\
$\xxb$&$0.94\pm0.27\pm0.15$&$2.38\pm0.30\pm0.21$\\
\hline\hline
\end{tabular}
\end{center}
\end{table}

According to the hadron helicity conservation, the angular
distribution of $\psipto B_8\bar{B_8}$ can be expressed as:
\begin{equation}
\frac{dN}{d\cos\theta} \propto 1+\alpha \cos^2\theta,
\end{equation}
where $\theta$ is the angle between $B_8$ and the beam
direction of the positron in the center-of-mass (CM) system.
In the
limit of infinitely heavy charm mass, hadron helicity
conservation implies $\alpha=1$~\cite{stanley} for both
$\jpsi$ and $\psip$ decays to octet baryon anti-baryon pairs.

Values of $\alpha$ for $\jpsi,~\psipto\ppb$ have been predicted
theoretically based on first order QCD. In the prediction of Claudson,
Glashow, and Wise~\cite{ang01}, the mass of the final baryon is taken
into account as a whole, while the constituent quarks inside the
baryon are taken as massless when computing the decay amplitude.  In
the prediction by Carimalo~\cite{ang02}, mass effects at the quark
level are taken into consideration.  Experimentally there are several
measurements for $\alpha$ for $\jpsito\ppb$, and the recent result of
$\alpha=0.676\pm0.036\pm0.042$ given by BES~\cite{lixl} is quite close
to Carimalo's prediction $\alpha=0.69$~\cite{ang02}. However, there is
only one measurement for $\psipto\ppb$, made by E835~\cite{e835}.
Results for $\psipto\ppb$ are summarized in Table~\ref{pre_alpha}. The
$\psipto\ppb$ events in BESII allow the measurement of $\alpha$, which
can be compared with the existing result and used to test hadron
helicity conservation.

\begin{table}[ht]
  \caption{\label{pre_alpha}Predicted and measured values of $\alpha$ for $\psipto\ppb$.}
  \begin{center}
    \begin{tabular}{ll} \hline\hline
      $\alpha$ value  & Source\\\hline
      Predicted value:&\\\hline
      $\alpha=0.58$   & Claudson \etal~\cite{ang01}\\
      $\alpha=0.80$   & Carimalo~\cite{ang02}\\\hline
      Measured value: &\\\hline
      $\alpha=0.67\pm0.15\pm0.04$& M. Ambrogiani \etal~\cite{e835}\\\hline\hline
    \end{tabular}
  \end{center}
\end{table}


BESII is a large solid-angle magnetic spectrometer which is
described in detail in Ref.~\cite{besii}. The momentum of  charged
particles is determined by a forty-layer cylindrical main drift
chamber (MDC) which has a resolution of
$\sigma_{p}$/p=$1.78\%\sqrt{1+p^2}$ ($p$ in GeV/$c$).  Particle
identification is accomplished using specific ionization ($dE/dx$)
measurements in the drift chamber and time-of-flight (TOF)
information in a barrel-like array of forty-eight scintillation
counters. The $dE/dx$ resolution is $\sigma_{dE/dx}\simeq8.0\%$;
the TOF resolution for Bhabha events is $\sigma_{TOF}= 180$ ps.
Radially outside of the time-of-flight counters is a
12-radiation-length barrel shower counter (BSC) comprised of gas
tubes interleaved with lead sheets. The BSC measures the energy
and direction of photons with resolutions of
$\sigma_{E}/E\simeq21\%/\sqrt{E}$ ($E$ in GeV),
$\sigma_{\phi}=7.9$ mrad, and $\sigma_{z}=2.3$ cm. The iron flux
return of the magnet is instrumented with three double layers of
proportional counters that are used to identify muons.

Monte Carlo (MC) simulation is used for mass resolution and
detection efficiency determination. In this analysis, a
GEANT3~\cite{geant} based MC package (SIMBES) with detailed
consideration of the detector performance (such as dead electronic
channels) is used. The consistency between data and MC has been
carefully checked in many high purity physics channels, and the
agreement is reasonable~\cite{simbes}.

%
The data samples used for this analysis consist of
$14.0\times10^6 (1\pm4\%)\ \psip$ events~\cite{psip} and
$6.42(1\pm4\%)$ pb$^{-1}$ of continuum data at
$\sqrt{s}=3.65$ GeV~\cite{cont}. The decay channels investigated
are $\psipto\ppb,~\llb,~\ssb$, and $\xxb$, where $\Lambda$ decays
to $p\pi^-$(63.9\%), $\Sigma^0$ decays to $\Lambda\gam$(100\%),
and $\Xi^-$ decays to $\Lambda\pi^-$(99.9\%).

\section{ \boldmath Event selection and branching ratio determination}
\subsection{\boldmath $\psipto\ppb$}
The experimental signature for the decay $\psipto\ppb$ is two
back-to-back, oppositely-charged tracks, each with a momentum of
1.586 GeV/$c$. The main backgrounds are: Bhabha and dimuon ($e^+ e^-
\to \mu^+ \mu^-$) events,
$\psipto\ee,~\mumu,~\pipi,~\kk$,
$\psipto\gam\chicj\ra\gam\pipi (\kk,~\ppb)$,
$\psipto\pi^0\pi^0\jpsi \ra \pi^0\pi^0\ee (\mumu)$,
$\psipto\pi^0\ppb$, etc.

The event selection requires two well reconstructed and oppositely
charged tracks. Each track is required to be well fitted to a
three dimensional helix, be in the polar angle region
$|\cos\theta|<0.8$, and have a momentum greater than 70 MeV/$c$ in
the xy-plane. The point of closest approach of each track to the
beamline is required to be within the interaction region which is
defined to be $\pm 20$ cm longitudinally and 2 cm radially.

In order to remove cosmic rays, the difference between the
time-of-flights of the positive and negative tracks,
$|t_+-t_-|$, is required to be less than 4.0 ns. Protons and
antiprotons are required to be identified by the TOF;
the measured time-of-flight of the track must be closest to the
prediction for the
proton/anti-proton hypothesis. Since $\psipto\ppb$ is a
back-to-back two-body decay, we require the acollinearity angle of
two tracks to be less than $5^o$. The deposited energy in the BSC of the
positive particle is required to be less than 0.75 GeV to remove possible
$\ee$ final state contamination. Finally, the energy sum
(calculated from the track's momentum) of the two tracks is required to be
within 130 MeV of the expected sum, 3.686 GeV, and the momentum
of the negative track is required to be within  150 MeV/$c$  of the
expected momentum 1.586 GeV/$c$.

\begin{figure}[h] \centering
\includegraphics[height=7cm]{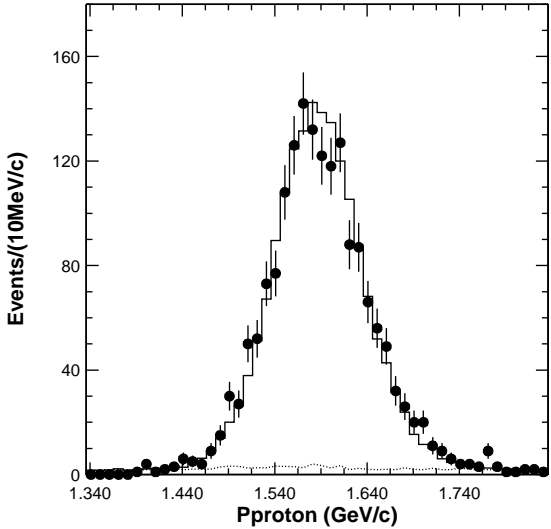}
\caption{ \label{ppb_fit} The fitted proton momentum spectrum. The
dots with error bars are data, the histogram is the fit to the
data including the signal shape from MC and all backgrounds,
and the dashed line is the background.}
\end{figure}

Events surviving the selection criteria are shown in
Fig.~\ref{ppb_fit} as dots with error bars. The same selection
criteria have been applied to background events
generated by the MC and normalized according to branching ratios listed in
PDG(2006)~\cite{pdg2006}, and 38.1 background events survive and are
shown as the dashed line in Fig.~\ref{ppb_fit}. The data are fitted
by a MC histogram for the signal plus a background function which
corresponds to the 38.1 simulated background events and a flat
distribution to describe the remaining background. From the
fit, the number of $\ppb$ events is determined to be $1618.2\pm43.4$,
where the error is statistical.



\subsubsection{ \boldmath Angular distribution of $\ppb$}
\begin{figure}[h] \centering
\includegraphics[height=7cm]{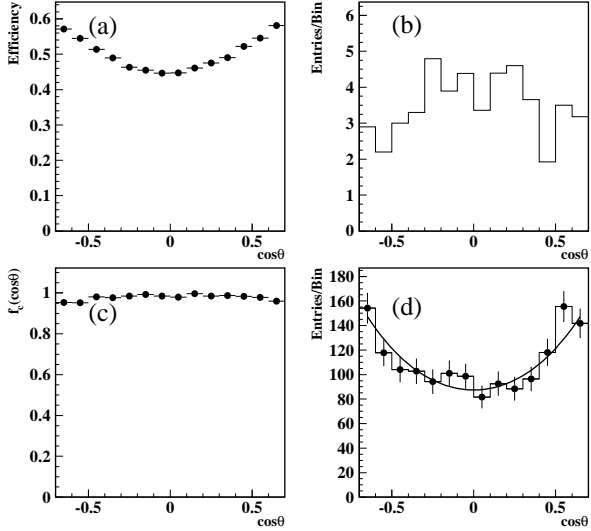}
\caption{ \label{alpha_fit} (a) The selection efficiency versus
$\cos\theta$ obtained from MC; (b) angular distribution of
background events from MC, which survive the same selection
criteria as used for data; (c) the correction obtained from data
($f_c(\cos \theta)$) to the MC efficiency; and (d) the angular
distribution of candidate $\psipto\ppb$ events.}
\end{figure}

To obtain the parameter $\alpha$ for $\psipto\ppb$, the
$\cos\theta$ dependence of the event selection efficiency must be
taken into account, which is determined using a flat angular
distribution ($\alpha$=0) in the MC simulation; see
Fig.~\ref{alpha_fit}(a). However, there are imperfections in the
MC simulation, which will distort the efficiencies determined by
the MC as a function of $\cos\theta$. In order to reduce this
systematic error, a correction to the MC efficiency is
made~\cite{lixl}. The correction factor $f_c(\cos\theta)$ is
defined as:
$$f_c(\cos\theta)=\frac{\varepsilon_{Data}}{\varepsilon_{MC}}(\cos\theta)=
\prod_{i}\frac{\varepsilon_{Data}(i)}{\varepsilon_{MC}(i)}(\cos\theta),$$
where $i$ denotes the selection criterion, $\varepsilon_{Data}(i)$
is the efficiency determined for data for criterion $i$, and
$\varepsilon_{MC}(i)$ is the efficiency from the MC for criterion
$i$.
The corrected MC efficiency is then:
$$\varepsilon^{\prime}_{MC}(\cos\theta)=\varepsilon_{MC}(\cos\theta)\times f_c(\cos\theta).$$

Due to the limitation on the number of $\psipto\ppb$ events, the
"reference" channel $\jpsito\ppb$ is chosen to determine the
correction factor due to its higher statistics and similar
kinematics. The selection criteria related to
the energy and momentum for $\psipto\ppb$ are scaled to the
reference channel $\jpsito\ppb$. Then following the re-weighting
procedures in Ref.~\cite{lixl} for our selection criteria, the
correction function $f_c(\cos\theta)$ is obtained and is shown in
Fig.~\ref{alpha_fit}(c).
With $\varepsilon_{MC}(\cos\theta)$ denoting the efficiency
obtained from $\psipto\ppb$ MC and $f_c(\cos\theta)$ the
correction function for the efficiency, we fit the measured angular
distribution of $\psipto\ppb$ data with the function
$N(\cos\theta)$,
$$N(\cos\theta)=N_0\times(1+\alpha
\cos^2\theta)\times\varepsilon_{MC}(\cos\theta)\times
f_c(\cos\theta),$$
as shown in Fig.~\ref{alpha_fit}(d).  The fit uses
a binned $\chi^2$ minimization method in the angular range $\cos\theta
\in$[-0.7,0.7] and gives $\chi^2_{min}=10.88$ for 12 degrees of
freedom.  The fitted value of the parameter $\alpha$ is $0.85\pm0.24$,
where the error is statistical.

As a consistency check, we also obtained $f_c(\cos\theta)$
directly from the $\psipto\ppb$ sample, and the fitted result
obtained using this correction yields $\alpha = 0.83 \pm 0.24$,
but its systematic uncertainty is 0.14, mainly due to the lower
statistics of the $\psipto\ppb$ sample, and much larger than the
systematic error of 0.04 determined using $f_c(\cos\theta)$
obtained from the $\jpsito\ppb$ sample (see section 3.1). This
demonstrates that $f_c(\cos\theta)$ determined from the
$\jpsito\ppb$ sample improves the systematic error on $\alpha$
without changing its central value and statistical error.

\subsubsection{ \boldmath Branching ratio of $\psipto\ppb$}
The selection efficiency is determined using $1\times10^5$
$\psipto\ppb$  MC events. The MC-determined efficiency is
$\epsilon_{MC}=(34.4\pm0.2)\%$, and the branching ratio is
determined to be:
$$Br(\psipto\ppb)=(3.36\pm0.09)\times10^{-4},$$
where the error is statistical.
\subsection{\boldmath $\psipto\llb$}
\begin{figure}[h] \centering
\includegraphics[height=7cm]{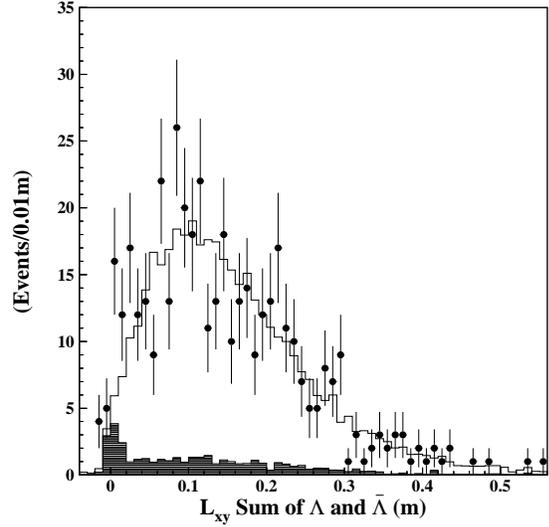}
\caption{\label{llb_lxy} The sum of the $\Lambda$ and
  $\lbar$ decay lengths. The histogram is the signal shape from the MC
  plus simulated background, the dots with error bars are data, and
  the shaded histogram is the background.}
\end{figure}

\begin{figure}[h] \centering
\includegraphics[height=7cm]{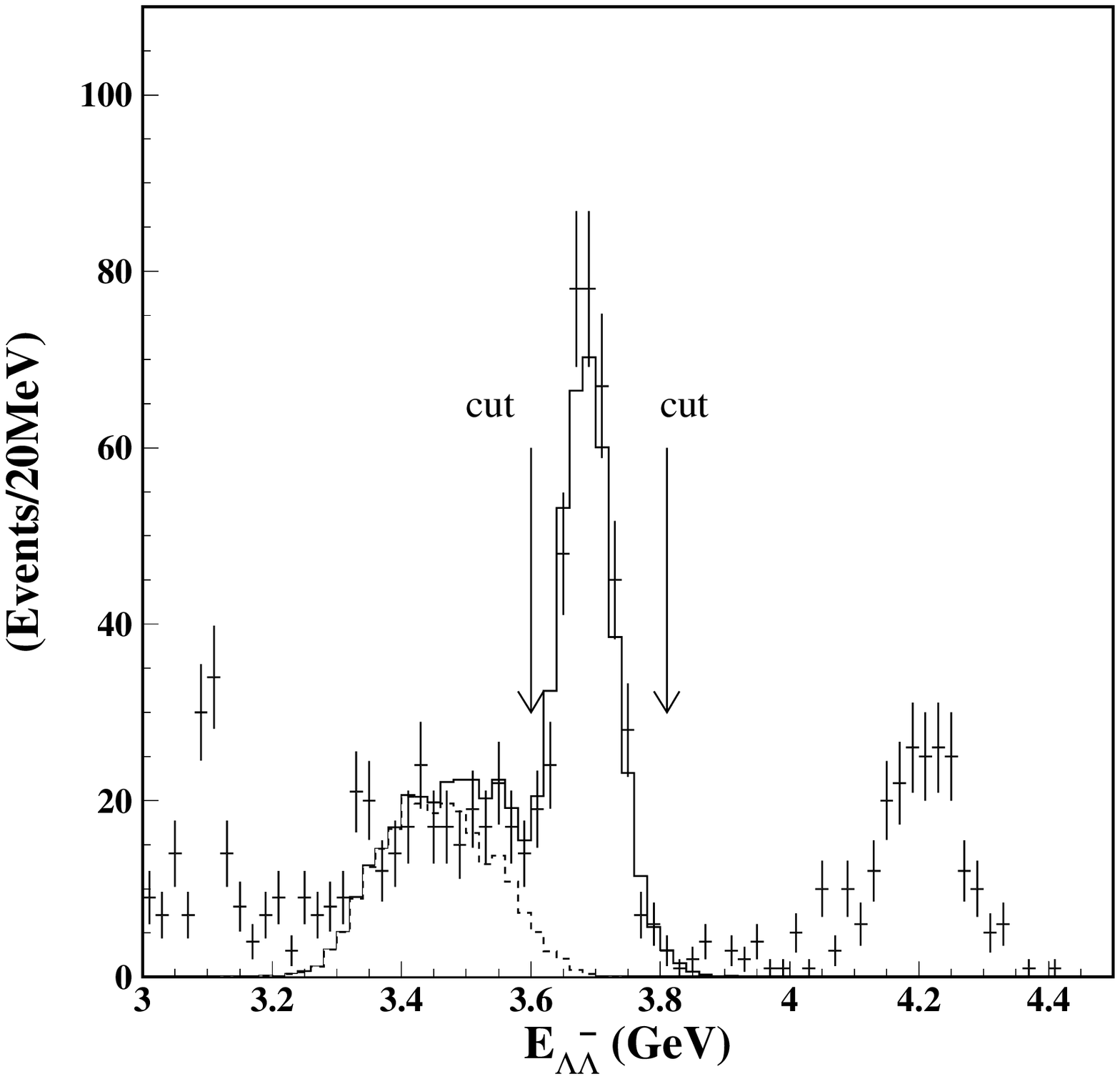}
\caption{\label{llb_ellb} The sum of the $\Lambda$ and $\lbar$ energies.
  The histogram is the signal shape from the MC plus backgrounds. The
  dots with error bars are data, and the dashed line is the main
  background from $\psipto\ssb$. The peaks at 3.1 and 4.2 GeV are
  from $\psipto\pi^0\pi^0\jpsi$, $\jpsito\llb$ and $\psipto\pipi\jpsi$,
  $\jpsito\ee$ (or $\mumu$), respectively.}
\end{figure}

Candidate events require four well reconstructed charged
tracks. The positive (negative) charged track with the higher momentum
is assumed to be the proton (antiproton); the other two are assumed
to be the $\pi^+$ and $\pi^-$. The two $p\pi$ pairs are required to pass
the $\Lambda$'s vertex finding algorithm successfully, and the
sum of the $\Lambda$ and $\lbar$ decay lengths must be greater
than 0.02 m (see Fig.~\ref{llb_lxy}). The sum of the $\Lambda$ and
$\bar{\Lambda}$ energies must be
in the region from 3.60 GeV to 3.81 GeV (see
Fig.~\ref{llb_ellb}). The missing momentum of the events should be
less than  0.18 GeV/$c$, and the difference between the measured mass
of $M_{\pbar\pi^+}$ and its expected value, $M_{\Lambda}$, should
be less than 12 Me/$c^2$ (three times the resolution of the
$M_{\Lambda}$).

\begin{figure}[h] \centering
\includegraphics[height=7cm]{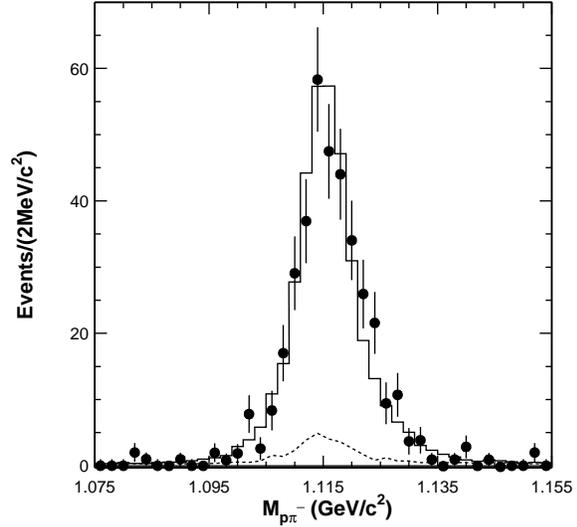}
\caption{ \label{llb_fit} The fitted $\Lambda$ mass spectrum.  The
dots with error bars are data, the histogram is the fit to
data which includes the signal shape from MC plus all backgrounds,
and the dashed line is the background.}
\end{figure}
The events that satisfy the selection criteria are shown in
Fig.~\ref{llb_fit} as dots with error bars; they are fitted by a
histogram of the signal shape from MC plus a background function
which describes the simulated backgrounds and a flat distribution
to describe any remaining sources. The simulated backgrounds are
mainly from $\psipto\ssb$ and $\psipto\Lambda\bar{\Sigma}^0+c.c.$
and normalizing according to branching ratios from PDG(2006), a
total of 32 background events are obtained.  The final number of
signal events from the fit is $337.2\pm19.9$.

The $\psipto\llb\ra\ppb\pipi$ efficiency is determined to be
$\epsilon_{MC}=(17.4\pm0.2)\%$ using $2\times10^5$ MC-simulated signal
events. The branching ratio is then:
$$Br(\psipto\llb)=(3.39\pm0.20)\times10^{-4},$$
where the error is statistical.
\subsection{\boldmath $\psipto\ssb$}
Candidate events are required to have four well reconstructed charged
tracks plus at least two good photons. The $\Lambda$ and $\lbar$
are selected using the method described in Section B. The
missing momentum of the events is required to be less than 0.25
GeV/$c$. The $\chi^2$ of the four constraint
(4C) kinematic fit to the hypothesis $\psipto\ppb\pipi\gam\gam$ must be less
than 20. The difference between the measured mass of
$M_{\pbar\pi^+\gam}$ and its expected value, $M_{\bar{\Sigma}^0}$,
should be less than 36 MeV/$c^2$ (three times the $M_{\bar{\Sigma}^0}$
resolution).

\begin{figure}[h] \centering
\includegraphics[height=7cm]{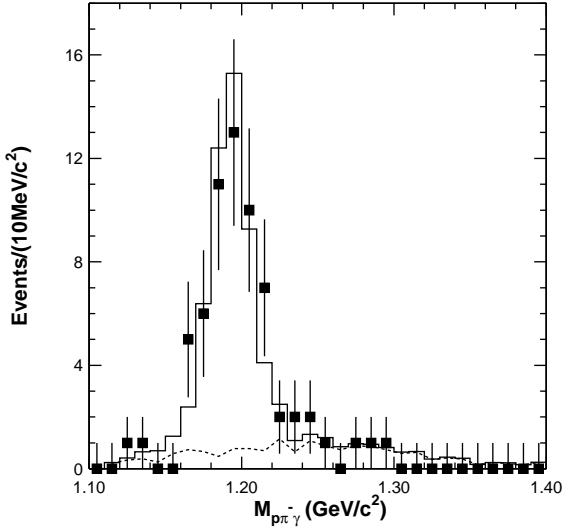}
\caption{ \label{ssb_fit} The fitted $\Sigma^0$ mass spectrum. The
dots with error bars are data, the histogram is the fit to
data which includes the signal shape from the MC and all backgrounds,
and the dashed line is the background.}
\end{figure}
The events that survive selection are shown in Fig.~\ref{ssb_fit} as
dots with error bars; they are fitted by a histogram of the signal
shape from MC plus a background function which describes the simulated
backgrounds and a flat distribution to describe the remaining
background. The main backgrounds are from $\psipto\llb$,
$\psipto\gam\chicj\ra\gam\llb$, $\psipto\Xi^0\bar{\Xi^0}$,
$\psipto\Lambda\bar{\Sigma^0}+c.c.$ and
$\psipto\Sigma^0\bar{\Xi^0}+c.c.$, and normalizing using branching
ratios from PDG(2006), 16.5 background events are
obtained.  The final number of signal events from the fit is
$59.1\pm9.1$.

The $\psipto\ssb\ra\llb\gam\gam\ra\ppb\pipi\gam\gam$ efficiency
is determined to  be  $\epsilon_{MC}=(4.4\pm0.1)\%$ using
$2\times10^5$ MC generated signal events.  The branching ratio of signal
channel is then:
$$Br(\psipto\ssb)=(2.35\pm0.36)\times10^{-4},$$
where the error is statistical.
\subsection{\boldmath $\psipto\xxb$}
\begin{figure}[h] \centering
\includegraphics[height=7cm]{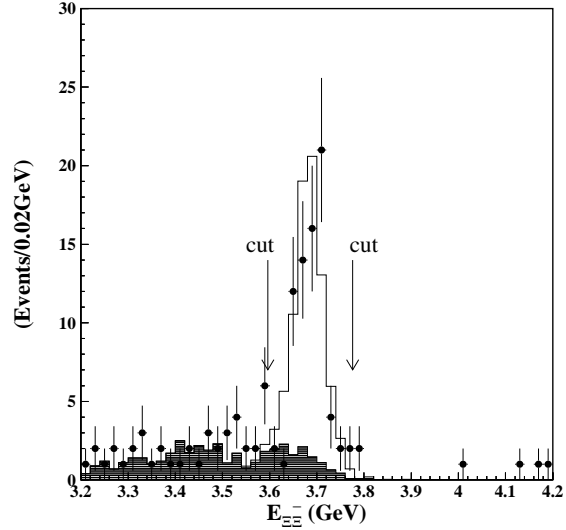}
\caption{\label{xxb_exxb} The $\Xi^-$ and
$\bar{\Xi}^+$ energy sum. The histogram is the signal shape from MC plus
background. The dots with error bars are data, and the
shaded area is the sum of simulated backgrounds.}
\end{figure}

Candidate events require six well reconstructed charged
tracks. The positive (negative) charged track with highest
momentum is assumed to be the proton (antiproton); the other four are
assumed to be $\pi$s. Looping over all possible $p\pi^-,~\pbar\pi^+$
combinations in an event, the one which successfully passes the
vertex finding algorithm and has the smallest value of
$\sqrt{(M_{p\pi^-}-M_{\Lambda})^2+(M_{\pbar\pi^+}-M_{\lbar})^2}$
is selected for further analysis. The energy sum of the $\Xi^-$ and
$\bar{\Xi}^+$ should be between 3.593 and 3.779  GeV (see
Fig.~\ref{xxb_exxb}), and the missing momentum of the events should be
less than 0.15 GeV/$c$. The difference between the measured mass of
$M_{\pbar\pi^+\pi^+}$ and its expected value, $M_{\bar{\Xi}^+}$,
should be less than 18 MeV/$c^2$ (three times the $M_{\Xi^-}$ resolution).

\begin{figure}[h] \centering
\includegraphics[height=7cm]{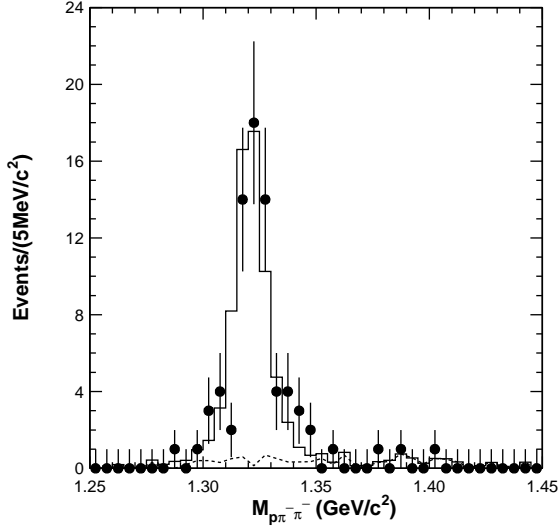}
\caption{ \label{xxb_fit} The fitted $\Xi^-$ mass spectrum.
Dots with error bars are data, the histogram is the fit to
data which includes the signal shape from the MC and all backgrounds,
and the dashed line is the sum of the backgrounds.}
\end{figure}
The events surviving selection are shown in Fig.~\ref{xxb_fit} as dots
with error bars, and they are fitted by a histogram of the signal
shape from MC plus a background function which describes the simulated
backgrounds and a flat distribution to describe remaining background.
The main background is from  $\psipto\pipi\jpsi\ra\pipi\llb$, and
normalizing by PDG(2006) branching fractions,
11.3 background events are obtained.
The final
number of signal events from the fit is $67.4\pm8.9$.

The  $\psipto\xxb\ra\llb\pipi\ra\ppb\pipi\pipi$ efficiency is
determined to be  $\epsilon_{MC}=(3.9\pm0.1)$\% using
$2\times10^5$ signal events generated by MC. The branching
ratio of the signal channel is then:
$$Br(\psipto\xxb)=(3.03\pm0.40)\times10^{-4},$$
where the error is statistical.

\section{ \boldmath Systematic error}
\subsection{\boldmath $\psipto\ppb$ angular distribution}

The systematic error on $\alpha$ in $\psipto\ppb$ decay from the
tracking reconstruction is determined using different MDC wire
resolution models in the MC simulations, which changes $\alpha$
by $2.7\%$. When the fit parameter of the efficiency
correction curve $f_c(\cos\theta)$ is changed by $1\sigma$,
$\alpha$ changes by $2.3$\%. The performance of the BES detector
has small differences between the time when the $58\times10^6$
$\jpsi$ events were obtained and when the $14\times10^6$ $\psip$
events were obtained. Using parameter files describing the
performance of BES detector at these two data taking periods, the
effect of this variation on $\alpha$ is determined to be $2.2$\%.
The effect of the background uncertainty on $\alpha$ is
negligible. Adding these contributions in quadrature gives a total
systematic error of $4.2$\%.
\subsection{Branching ratios}
The systematic errors on the branching ratios are mainly from the
uncertainties in the MDC tracking, $\alpha$, the
hadronic interaction model, background estimations, and differences
between data and MC for the
$\Lambda$ vertex finding, decay length requirement, and kinematic fitting.

The MDC tracking gives a systematic error of about $2$\% for a
proton or anti-proton~\cite{simbes} and $1$\% for a low-energy
$\pi$, which is determined from the channel
$\psipto\pipi\jpsi\ra\pipi\mumu$. The detection efficiency depends
on the angular distribution of the baryon pair. For $\ppb$ decay,
when changing the $\alpha$ value by $1\sigma$, the branching ratio
changes by $2.4$\%; in the other three channels, $\alpha = 0.5$ is
used as a nominal value, the maximum differences for
$|\epsilon_{\alpha=0.5}-\epsilon_{\alpha=0}|$ and
$|\epsilon_{\alpha=0.5}-\epsilon_{\alpha=1}|$ are taken as
systematic errors, they are $6.5$\%, $7.6$\%, $6.8$\%,
respectively. The uncertainties of the detection efficiencies
caused by assumed flat angular distributions for secondary decay
of baryons are much smaller than those from angular distributions
of $\psip$ to baryon pair primary decays, and are therefore
neglected here~\cite{pingrg}. Different simulation models for the
hadronic interaction (GCALOR/Geant-FLUKA)~\cite{gcalor,fluka} give
different efficiencies, giving systematic errors of $2.18$\%,
$0.46$\%, $0.00$\%, $1.08$\% for the studied channels,
respectively. The background uncertainty is studied by changing
the nominal branching ratios of the backgrounds which have large
statistical errors. If the branching ratios of the background
channels are changed by $100$\% in the $\ppb$, $\llb$, and $\xxb$
channels, the changes in the branching ratios in the signal
channels are $0.1$\%, $1.0$\% and $0.2$\%, respectively. For the
$\ssb$ channel, where the shape of the simulated backgrounds is in
good agreement with the data in the $\Lambda\gam$ invariant mass
distribution, the branching ratios of backgrounds are only changed
by $20$\%, resulting in a change of the branching ratios of the
signal channel of $2.3$\%. According to the reference channel
$\jpsito\llb$~\cite{bianjg}, the secondary vertex finding of
$\Lambda$ gives a systematic error of $0.7$\% for each $\Lambda$
vertex, and the requirement on the sum of the decay length
contributes $1.4$\%.

In the branching ratio determination of four channels, the
continuum contribution must be subtracted. The continuum data are
also selected with the same criteria as for the $\psip$ decay
signal channels, and the number of the surviving events times a
luminosity normalization factor is taken as a systematic error.
The kinematic fit of $\ppb\pipi\gam\gam$ in $\psipto\ssb$ gives a
systematic error of $7.6$\% from the reference channel
$\psipto\pipi\jpsi\ra\pipi\pipi\pi^0$~\cite{3pi}. The uncertainty
on the total number of $\psip$ events is $4$\%. The systematic
errors of the acollinearity angle, $E_{B\bar{B}}$ region, baryon
mass (or momentum), and $P_{miss}$ requirements are studied with
corresponding $\jpsi \to p \bar{p}$ decays.

In the $\psipto\ppb$ selection, the systematic errors due to the
uncertainties from particle identification, the cosmic ray veto, and
the deposited energy criterion are studied by this channel itself.
All the systematic errors in the branching ratio measurements are
summarized in Table~\ref{tab_sys}.

\section{ \boldmath Summary and discussion}
Based on $14\times10^6$ $\psip$ events, the branching ratios of
$\psipto\ppb$, $\llb$, $\ssb$, and $\xxb$ are measured, the
results are listed in Table~\ref{result}, together with the ratios
of $\psipto B\bar{B}$ to $\jpsito B\bar{B}$. They are in agreement
with the results published by the CLEO collaboration~\cite{cleo}
within $2\sigma$ for $\ppb$ and within $1\sigma$ for the other
three channels. The differences of the branching fractions between
current measurements and those of BESI are $2.5\sigma$,
$3.1\sigma$, $1.5\sigma$, $3.5\sigma$ for the four channels,
respectively. 

The angular distribution parameter $\alpha$ for $\psipto\ppb$ is
measured to be $0.85\pm0.24\pm0.04$, which is in agreement within
$1\sigma$ with the E835 result~\cite{e835}, and close to
Carimalo's prediction~\cite{ang02}.

\begin{table}[h!]
\caption{\label{tab_sys}Systematic errors in the branching ratio
measurements (\%).}
\begin{center}
\begin{tabular}{lcccc}\hline\hline
Source & $\ppb$ & $\llb$ & $\ssb$ & $\xxb$ \\\hline
MDC tracking & 4 & 4.5 & 4.5 & 5.7 \\
PID & 2.4 & & & \\
Cosmic Ray Exc. & 0.9 & & & \\
Deposit Energy  & 0.9 & & & \\
Acol. angle      & 0.9 & & & \\
Vtx. finding & & 1.4 & 1.4 & 1.4 \\
Decay length & & 1.0 & 1.0 & \\
$E_{B\bar{B}}$, $M_B$ (or $P_B$) & 0.8 & 0.6 & 1.6 & 1.6 \\
$P_{miss}$ & & 1.6 & 0.5 & 1.7 \\
$\gam$ tracking & & & 4 & \\
Kinematic fit & & & 7.6 & \\
Bg. Esti. & & 1.0 & 2.3 & 0.2 \\
Continuum data & 0.8 & 1.0 & & \\
$\alpha$ value & 2.4 & 6.5 & 7.6 & 6.8 \\
Hadronic Interaction & 2.2 & 0.5 & & 1.1 \\
$N_{\psip}$  & 4 & 4 & 4 & 4 \\\hline Total error & 7.3 & 9.4 &
13.4 & 10.3 \\\hline\hline
\end{tabular}
\end{center}
\end{table}

\vspace*{-0.5mm}

\begin{table}[h!]
  \caption{\label{result}{Branching ratios of $\psip$ decays into baryon
  anti-baryon pairs. The first error is statistical and the second
  systematic. The value Q is $BR(\psipto B\bar{B})/BR(\jpsito B\bar{B})$.
  The  $\jpsi$ branching ratios are taken from Ref.~\cite{lixl} for
  $\ppb$, Ref.~\cite{bianjg} for $\llb$ and $\ssb$, and Ref.~\cite{pdg2006}
  for $\xxb$.}}
\begin{center}
\begin{tabular}{lll}\hline\hline
modes& BRs ($\times10^{-4}$)& Q (\%)\\
\hline
$\ppb$&$3.36\pm0.09\pm0.25$& $14.9\pm1.4$\\
$\llb$&$3.39\pm0.20\pm0.32$& $16.7\pm2.1$\\
$\ssb$&$2.35\pm0.36\pm0.32$& $16.8\pm3.6$\\
$\xxb$&$3.03\pm0.40\pm0.32$& $16.8\pm4.7$\\
\hline\hline
\end{tabular}
\end{center}
\end{table}

\section{ \boldmath Acknowledgement}
The BES collaboration thanks the staff of BEPC for their hard
efforts. This work is supported in part by the National Natural
Science Foundation of China under contracts Nos. 10491300,
10225524, 10225525, 10425523, the Chinese Academy of Sciences
under contract No. KJ 95T-03, the 100 Talents Program of CAS under
Contract Nos. U-11, U-24, U-25, and the Knowledge Innovation
Project of CAS under Contract Nos. U-602, U-34 (IHEP), the
National Natural Science Foundation of China under Contract No.
10225522 (Tsinghua University), and the Department of Energy under
Contract No.DE-FG02-04ER41291 (U Hawaii).

\end{document}